\documentclass[%
aps,prx,letter,preprintnumbers,superscriptaddress,nofootinbib,longbibliography,floatfix
]{revtex4-2}
\usepackage{url}

\usepackage{graphicx}  
\usepackage{dcolumn}   
\usepackage{bm}        
\usepackage{amssymb}   
\usepackage{slashed}
\usepackage{xcolor}
\usepackage{array}
\usepackage{caption}
\usepackage{lineno}
\usepackage{subcaption}
\expandafter\let\csname equation*\endcsname\relax
\expandafter\let\csname endequation*\endcsname\relax
\usepackage{amsmath}
\usepackage{booktabs}
\usepackage{multirow}
\usepackage{subcaption}
\usepackage{wrapfig}
\usepackage{dblfloatfix}
\usepackage{url}
\usepackage[shortlabels]{enumitem}
\usepackage{parskip}

\usepackage{hyperref}
\hypersetup{
  colorlinks=true,
  citecolor=blue,
  linkcolor=blue,
  urlcolor=blue
}

\begin{document}

\title{Advancing Set-Conditional Set Generation: \\Diffusion Models for Fast Simulation of Reconstructed Particles}

\author{Dmitrii Kobylianskii}
\email{dmitry.kobylyansky@weizmann.ac.il}
\affiliation{Weizmann Institute of Science, Israel}
\affiliation{These authors contributed equally}

\author{Nathalie Soybelman}
\email{nathalie.soybelman@weizmann.ac.il}
\affiliation{Weizmann Institute of Science, Israel}
\affiliation{These authors contributed equally}

\author{Nilotpal Kakati}
\affiliation{Weizmann Institute of Science, Israel}

\author{Etienne Dreyer}
\affiliation{Weizmann Institute of Science, Israel}

\author{Benjamin Nachman}
\affiliation{Lawrence Berkeley National Laboratory, Berkeley, USA} 

\author{Eilam Gross}
\affiliation{Weizmann Institute of Science, Israel}

\date{\today}

\begin{abstract}
The computational intensity of detector simulation and event reconstruction poses a significant difficulty for data analysis in collider experiments. This challenge inspires the continued development of machine learning techniques to serve as efficient surrogate models. We propose a fast emulation approach that combines simulation and reconstruction. In other words, a neural network generates a set of reconstructed objects conditioned on input particle sets. To make this possible, we advance set-conditional set generation with diffusion models. Using a realistic, generic, and public detector simulation and reconstruction package (COCOA), we show how diffusion models can accurately model the complex spectrum of reconstructed particles inside jets.
\end{abstract}
\maketitle

\section{Introduction}
For experiments at the Large Hadron Collider (LHC), the demand for simulated data has increased throughout recent data-taking periods and will surge by roughly a factor of ten during the High-Luminosity runs \cite{hl-lhc}.
The standard simulation pipeline used at the general-purpose experiments ATLAS and CMS is designed to closely mirror that of real data \cite{SOFT-2010-01, Hildreth:2015kps}. Initially, proton-proton collision events are generated, followed by parton shower, hadronization, and secondary decays. The set of stable, or ``truth'', particles then enter the detector volume, and their material interactions are modeled on a microscopic level using \textsc{Geant4}~\cite{geant,geant2,geant3}. The resulting energy deposits in the tracker and calorimeter cells seed the formation of tracks and clusters, respectively.
Subsequently, a reconstruction algorithm refines these data into objects suitable for statistical analysis. 
Depending on the algorithm, the reconstructed ``particle flow objects'' (PFOs) either represent particles directly or else consist of tracks and calorimeter clusters with kinematics adjusted to maintain energy flow at the jet level.

From the perspective of computational efficiency, the main constraint in the traditional simulation pipeline comes from the highly sophisticated, but expensive model of particle-detector interactions provided by \textsc{Geant4}. In a bid to remove this constraint, various machine learning approaches are currently being explored~\cite{Paganini:2017hrr,detsimoverview}. These rapid surrogate models aim to drastically reduce computational time while otherwise performing similarly to the traditional simulation pipeline. 

The majority of research efforts in this direction have concentrated on fast calorimeter simulation~\cite{Paganini:2017hrr,detsimoverview, detsimdiff,Butter:2022rso}. In this approach, the objective is to replace the slowest link in the simulation chain while preserving the other steps, including reconstruction. Utilizing the features of an incoming particle as a conditioning factor, a generative model is employed to produce the detector response, which can then be aggregated for all particles within an event~\cite{atlasfast3}. Within the scope of the Fast Calorimeter Challenge~\cite{ds1,ds2,ds3}, a wide array of methods have been developed and analyzed. Various architectures, including those based on Generative Adversarial Networks (GANs)~\cite{caloshowerGAN,deeptreeGAN}, normalizing flows~\cite{caloflow, caloINN} and diffusion models~\cite{Mikuni:2022xry,CaloDiffusion,caloclouds,caloscore,calograph} have been introduced. These approaches represent the calorimeter's response through diverse formats such as images, point clouds, or graphs.

A complementary approach is to replace the entire simulation pipeline by combining event generation, detector simulation, and event reconstruction. While reconstruction has typically not been a computational bottleneck, this is changing with improvements in fast detector simulations and with increased combinatorics from higher instantaneous luminosities~\cite{CERN-LHCC-2022-005,Software:2815292}. While we are not aware of a model demonstrated for entire events at the level of individual PFOs, there are a number of proposals using high-level objects~\cite{Hashemi:2019fkn}. 
%
This approach is highly efficient as it encompasses all simulation steps in a single process, eliminating the need for additional processing. However, it is worth noting that this method is process-dependent, necessitating optimization and retraining of the network for each analysis. 
A related strategy entails creating a generative model for individual jets, the most complex part of entire events~\cite{epicgan, jetben, jetepic, pcdroid, MPGAN, endtoend_flow,Birk:2024knn,Mikuni:2024qsr}. In this scenario, the constituents of the jet are generated conditioned on both the jet features and the particle type. This approach is less process-dependent than learning entire events, but it still requires the model to learn about the physics of jet substructure and when moving to entire events, it necesitates event-splitting and additional processing for other objects within the event. 

Our goal is to develop a fast and automated end-to-end strategy that starts at truth particles and ends with reconstructed particles. By only learning the detector response and not the full event generation, we strive to be process independent. Going all the way to PFOs addresses the computational bottleneck from reconstruction and keeps the dimensionality of the problem manageable. 
In the pursuit of a process-independent, full-event method, a novel strategy for fast simulation known as \textit{FlashSim}~\cite{flashsim,Vaselli:2024vrx} has been introduced\footnote{The model presented in Ref.~\cite{Howard:2021pos} has similar goals, but there is no matching between the truth and reconstructed objects during training.}. This method involves utilizing stable particles to predict high-level objects such as jets, fat jets, muons, electrons, and more. For each object, a separate network is trained using distinct input and target variables.
Extending the end-to-end methodology to indiviudal truth and reconstructed particles requires going beyond all examples mentioned above because we need to generate a variable-length set of particles conditioned on another variable-length set of particles. A prototype for this approach was presented in Ref.~\cite{fastsim} using a simplified version of the problem. This initial model solely considered charged particles within a jet, with smeared tracks as targets. The architecture incorporated graph neural networks with slot attention mechanisms. Now, transitioning towards a more realistic approach that encompasses both neutral and charged particles while targeting reconstructed particles, we explore diffusion models, showcasing enhanced performance compared to the baseline slot attention model. These models are studied using a detailed and realistic detector simulation and reconstruction. We will show that continuous normalizing flows~\cite{chen2019neural} are the most effective approach of the methods we studied. This forms the bassis of a companion paper~\cite{letter} that shows how an end-to-end fast simulation of the CMS experiment can outperform classical baselines.

This paper is organized as follows. Section~\ref{sec:data} introduces the dataset, with both the detailed simulation and reconstruction steps. The neural network approaches are described in Sec.~\ref{sec:algorithms} and trained in Sec.~\ref{sec:training}. Numerical results are presented in Sec.~\ref{sec:results} and the paper ends with conclusions and outlook in Sec.~\ref{sec:conclusions}.

\section{Dataset}
\label{sec:data}
The network takes as input a single jet of truth particles entering the detector\footnote{We consider single jets here as the most complex objects; future work will explore entire events.}. For the generation task, the target is the set of reconstructed particles. Instead of relying on a parametrized smearing model (e.g. \textsc{Delphes}~\cite{delphes}), we obtain reconstructed particles using a realistic detector simulation followed by a particle flow algorithm described below. Each input or output reconstructed particle is represented by its momentum, direction, and charge ($p_\text{T}$, $\eta$, $\phi$, $|q|$). Notably, charge prediction is a new feature introduced in this study, as our previous work utilized a toy model focusing solely on charged particles. Generally, reconstructed particles are classified into five classes: charged hadrons, neutral hadrons, photons, electrons, and muons. Given the significant class imbalance within the dataset, we simplify the task by predicting only whether the object is neutral or charged.

\subsection{Truth Event Generation}
We concentrate on a localized reconstruction of particles within a single jet. For this purpose, we utilize \textsc{Pythia8.3}~\cite{pythia} to generate a single quark with momentum ranging between 10 GeV and 200 GeV, with the initial direction randomly selected within the ranges $|\eta| < 2.5$ and $-\pi < \phi < \pi$. Subsequent to the parton shower and hadronization, only stable particles with momentum exceeding 1 GeV and $|\eta| < 3.0$ are retained. The set of particles contained within the jets exhibits an average cardinality of $N = 6.7$, with a maximum of 25 particles and a minimum of 1 particle per truth event, as measured on the whole dataset. 

\subsection{Detector simulation}
The detector simulation uses \textsc{COCOA}~\cite{cocoa}, a \textsc{Geant4}-based configurable calorimeter simulation toolkit. The detector comprises three electromagnetic and three hadronic calorimeter layers, simulating ATLAS-like materials. Geometrically, the coverage is divided into a barrel region ($|\eta| < 1.5$) and two endcaps ($1.5 < |\eta| < 3.0$). The cell depth follows a $1/\cosh{\eta}$ profile to maintain a constant effective interaction depth across $\eta$. Moreover, the tracking region of the detector is subjected to a magnetic field of 3.8~T. To avoid particles created upstream of the calorimeter, our detector simulation is simplified by assuming that the tracker contains no material.

Tracking effects are emulated by first computing the trajectories of charged particles within the magnetic field, followed by applying smearing to the track parameters $q/p$, $\theta$, and $\phi$. Additionally, we discard tracks originating far from the beamline, with a transverse radius exceeding $75$ mm ($250$ mm) in the barrel (endcap) region.

Following the simulation, a topological clustering algorithm is employed to group cells based on their deposited energy, expected noise levels, and proximity. These calorimeter clusters, along with the cells belonging to them and the set of tracks, are the input for the reconstruction algorithm.
\subsection{Particle Reconstruction}\label{sec:reconstruction}

To reconstruct particles, we utilize a recently developed machine learning algorithm called HGPflow~\cite{HGpflow}. The design principle of HGPflow is to learn the energy assignment between the input set of tracks and calorimeter clusters and an output set of particles using a hypergraph prediction network~\cite{hypergraph}. We trained a version of HGPflow on a statistically independent dataset of 294k single jets and evaluated this model to obtain a set of reconstructed particles for each jet in our training dataset. These predictions serve as the target for our fast simulation models.
\subsection{Preprocessing}
To mitigate contamination from detector noise potentially converted into neutral particles during HGPflow reconstruction, we employ a filtering process. Neutral reconstructed particles located outside a cone of $R=0.4$ around the truth-jet axis are removed since these typically correspond to clusters of calorimeter noise. Subsequently, before inputting the object features into the networks, we apply relative scaling. For each event individually, we scale $\log p_T$, $\eta$, and $\phi$ based on the mean and standard deviations of the truth particles. The resulting scaling quantities of $\eta$ and $\log p_T$ are then added as global features to retain knowledge of the absolute jet properties.
Additionally, we remove events where truth particles have a $|\phi| > 2.8$ to avoid edge effects.
\section{Fast simulation algorithms}
\label{sec:algorithms}
\begin{figure}[t]
    \centering
    \includegraphics[width=0.85\textwidth]{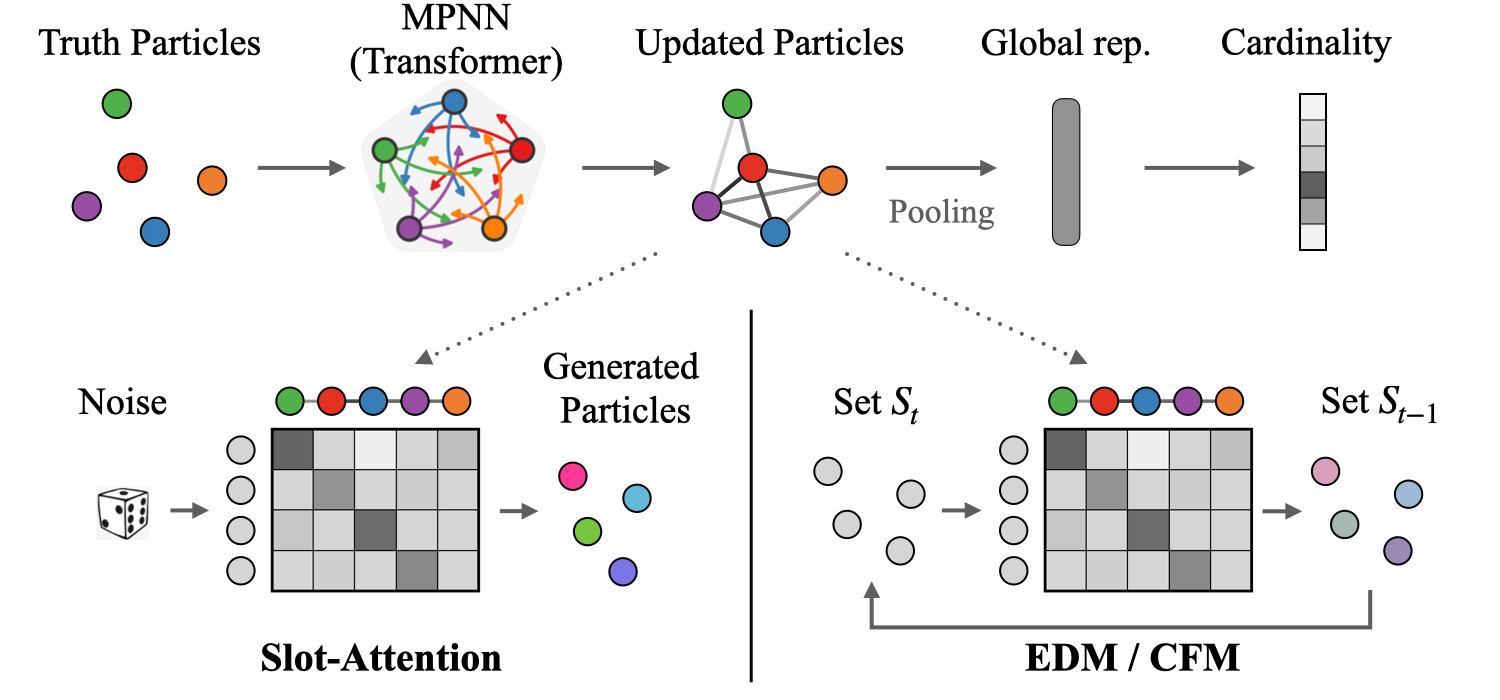}
    \caption{Simplified sketches of the two proposed network architectures for the set generation task. EDM and CFM extend the Slot Attention approach presented in detail in~\cite{fastsim} with an iterative denoising procedure.}
    \label{fig:all_nets}
\end{figure}

In this study, we evaluate the performance of two set-conditional set generation approaches. Both methods rely on graph neural networks, as they can naturally represent unordered, set-valued data through a graph structure. Currently, we use fully connected graphs because we have single jets. This can be readily generalized when considering full events where different event regions are loosely or not at all connected. Edges within the graph allow inter-particle relations to be encoded through message passing. Simplified illustrations of the network architectures are depicted in Fig.~\ref{fig:all_nets}. 
\subsection{Slot Attention (SA)}
The slot-attention approach, initially proposed in Ref.~\cite{fastsim}, constitutes the baseline model. In this approach, the generation task is divided into two stages: cardinality and feature prediction. Cardinality is inferred from the updated representations of truth particles following message passing. Subsequently, based on the predicted cardinality, particles are initialized with random noise and iteratively refined through a slot-attention block~\cite{slotattention}, which attends to the updated truth particle representations. The introduction of noise during initialization is crucial to prevent deterministic predictions. A WGAN~\cite{arjovsky2017wasserstein}-like setup is used to train the model whereby a distribution-level distance is used to steer the generator. In our previous and current work, we recognize(d) that the learned noise model exhibits room for improvement, indicated by the limited precision of the predictions.

We explore diffusion models in order to address these limitations. Diffusion models are state-of-the-art in generative AI and aim to directly learn explicitly or implicity the probability density (or the gradient of the density). Details of the EDM Diffusion (EDM) and Continuous Flow Matching (CFM) approaches are detailed in the subsequent section. The CFM is based on normalizing flows, but is still a variation on diffusion models; we will refer to both EDM and CFM as `diffusion' while EDM will refer specifically to the score-based approach described below.

\subsection{Diffusion Models}
In this approach, we maintain the cardinality prediction method as originally introduced in the slot-attention approach and solely modify the feature prediction task. After noting that a single slot-attention block lacks the expressiveness needed to transform complete noise into sufficiently accurate features, we opt for a different approach. This involves replacing the feature prediction component with a diffusion process or flow matching. Starting with a fully noised set of particles, we use a transformer-based network to iteratively predict a feature estimate in the case of diffusion or a vector field in the case of the flow matching objective. This output is used to update the noised set until we achieve the final, denoised reconstructed particles.
\begin{table}[b]
    \centering
    \begin{tabular*}{0.8\linewidth}{@{\extracolsep{\fill}}lr}
        \toprule
            \multicolumn{2}{@{}l}{\textbf{Network and preconditioning}}\\
            \addlinespace[2mm]
            Skip scaling $c_{\text{skip}}(\sigma)$ & $\sigma_{\text{data}}^2/(\sigma^2 + \sigma_{\text{data}}^2)$\\
            \addlinespace[2mm]
            Output scaling $c_{\text{out}}(\sigma)$ & $\sigma \cdot \sigma_{\text{data}}/\sqrt{\sigma^2 + \sigma_{\text{data}}^2}$\\
            \addlinespace[2mm]
            Input scaling $c_{\text{in}}(\sigma)$ & $1/\sqrt{\sigma^2 + \sigma_{\text{data}}^2}$ \\
        \midrule
            \multicolumn{2}{@{}l}{\textbf{Training}}\\
            \addlinespace[2mm]
            Loss weighting $\lambda(\sigma)$ & ($\sigma^2 + \sigma_{\text{data}}^2)/(\sigma \cdot \sigma_{\text{data}})^2$\\
        \midrule
            \multicolumn{2}{@{}l}{\textbf{Sampling}}\\
            \addlinespace[2mm]
            ODE solver & 4-th order PNDM~\cite{liu_pseudo_2022} method\\
            \addlinespace[2mm]
            Time steps & $(\sigma_\text{max}^{1/\rho} + \frac{i}{N - 1}(\sigma_{\text{min}}^{1/\rho} - \sigma_{\text{max}}^{{1/\rho}}))^\rho$\\
        \midrule
            \textbf{Parameters} & $\sigma_{\text{min}} = 0.002$, $\sigma_{\text{max}} = 80$, $\sigma_{\text{data}} = 1.1$, $\rho = 7$\\
        \bottomrule
    \end{tabular*}
    \caption{Parameters of the EDM diffusion model formulation.}
    \label{tab:edm_spec}
\end{table}

\subsubsection{EDM Formulation}

We adopt score-based diffusion, a method where a diffusion process gradually perturbs the original data $x$ with Gaussian noise while the neural network learns the time-dependent score function $\nabla_x \log p(x; \sigma)$. This function enables us to reverse the noise process, starting from pure noise $x_T \sim \mathcal{N}(0, 1)$ and iteratively denoising it to sample from the original data distribution $x_0 \sim p_\text{data}$.

We adapt the EDM strategy introduced in Ref.~\cite{karras_2022} for score-based denoising diffusion models. During training, we sample noise rates $\sigma$ from the log-normal distribution, defined by $\log(\sigma) \sim \mathcal{N}(-0.8, 0.8)$. The network receives scaled noisy data $x = c_{in}(\sigma)(x_0 + \sigma \epsilon)$ as input, where $\epsilon \sim \mathcal{N}(0, 1)$. To enhance network predictions, we combine the neural network output $F_{\theta}(x; \frac{1}{4}\ln{\sigma}, c)$ with a skip connection $D_\theta(x; \sigma, c) = c_{skip}(\sigma)x + c_{out}(\sigma)F_\theta(x;\frac{1}{4}\ln{\sigma}, c)$, where $c$ denotes contextual information.
The weighted loss function minimized during training is then defined as: 
\begin{equation}
    \mathcal{L} = E_{\sigma, x_0, \epsilon, c} \big[\lambda(\sigma)\left\|D_\theta(x; \sigma, c) - x_0 \right\|^2\big]\,.
    \label{eq:diffloss}
\end{equation}
We adopted specific design choices from Ref.~\cite{karras_2022}, as detailed in Tab.~\ref{tab:edm_spec}. During the inference process, we utilize the network predictions to compute the score function $\nabla_x \log p(x; \sigma, c) = (D_\theta(x; \sigma, c) - x)/\sigma$ and solve the diffusion ordinary differential equation (ODE) using PNDM method from Liu et al.~\cite{liu_pseudo_2022}.
\subsubsection{Flow Matching Formulation}
Continuous normalizing flows (CNF)~\cite{chen2019neural} are a type of generative model that constructs a transformation from samples $x_0$ of a base distribution $p_0$ to samples $x_1$ of a target distribution $p_1$ using an ordinary differential equation (ODE):
\begin{equation}
    \mathrm{d}x = v_{\theta}(t, x) \mathrm{d}t,
\end{equation}
where the vector field $v_{\theta}$ is represented by a neural network and $t$ ranges from 0 to 1. Lipman et al.~\cite{lipman_flow_2023} recently proposed a Flow Matching (FM) objective, which allows the vector field to be learned through a regression task:
\begin{equation}
    \mathcal{L}_{\text{FM}}(\theta) = \mathbb{E}_{t,p_t(x)}||v_{\theta}(t, x) - u_t(x)||^2\,.
\end{equation}
The function $u_t$ defines the desired probability path between $p_0$ and $p_1$, though it is generally not known in a closed form. This issue can be addressed with the Conditional Flow Matching (CFM) objective~\cite{lipman_flow_2023,tong2024improving,liu_flow_2022}, optimizing which is equivalent to optimizing the FM objective. In this formulation, $u_t$ and the probability path $p_t$ are built in a sample-conditional manner:
\begin{equation}
    \mathcal{L}_{\text{CFM}}(\theta) = \mathbb{E}_{t, q(z),p_t(x|z)}||v_{\theta}(t, x) - u_t(x|z)||^2,
    \label{eq:cfm_loss}
\end{equation}
where $t \in [0, 1]$, $z \sim q(z)$, and $x \sim p_t(x | z)$. Our approach uses the modified CFM framework from~\cite{lipman_flow_2023}. By identifying the condition $z$ with the single sample $x_1$ (in our context, a set of PFOs), we define:
\begin{equation}
    \begin{aligned}
        p_t(x|z) &= \mathcal{N}(x | tx_1, (t\sigma-t+1)^2)\,,\\
        u_t(x|z) &= \frac{x_1 - (1 - \sigma)x}{1 - (1-\sigma)t}\,,
    \end{aligned}
    \label{eq:prob_path}
\end{equation}
which outlines a linear trajectory between a standard normal distribution and a Gaussian distribution centred at $x_1$ with a standard deviation of $\sigma$, chosen as $10^{-4}$.

Following the strategy in Ref.~\cite{dax_flow_2023}, we sample $t$ from a power-law distribution $p(t) \propto t^{\frac{1}{1 + \alpha}}, \ t \in [0, 1]$. Setting $\alpha = 0$ results in a uniform distribution, while $\alpha > 0$ gives more weight to probability paths corresponding to larger $t$ values. Our experiments showed that $\alpha = 2$ achieves the best results.

\subsubsection{Architecture Description}
The EDM and CFM models share the same transformer-based architecture; network hyperparameters are shown in~\autoref{tab:hyperparams}. To improve sampling quality, we incorporate self-conditioning on the network output from the previous timestep, following a technique proposed in Ref.~\cite{chen_2023}. This involves concatenating $x_t$ with the previously estimated $\tilde{x}/\tilde{v}$. For training the network in this context, we set $\tilde{x}/\tilde{v} = 0$ with a probability of $p = 0.5$ reverting to the model without self-conditioning. The remaining time, first estimate $\tilde{x}/\tilde{v}$ and then utilize it for the self-conditioning with detached gradients.

\begin{figure}[t]
    \centering
    \includegraphics[width=\textwidth]{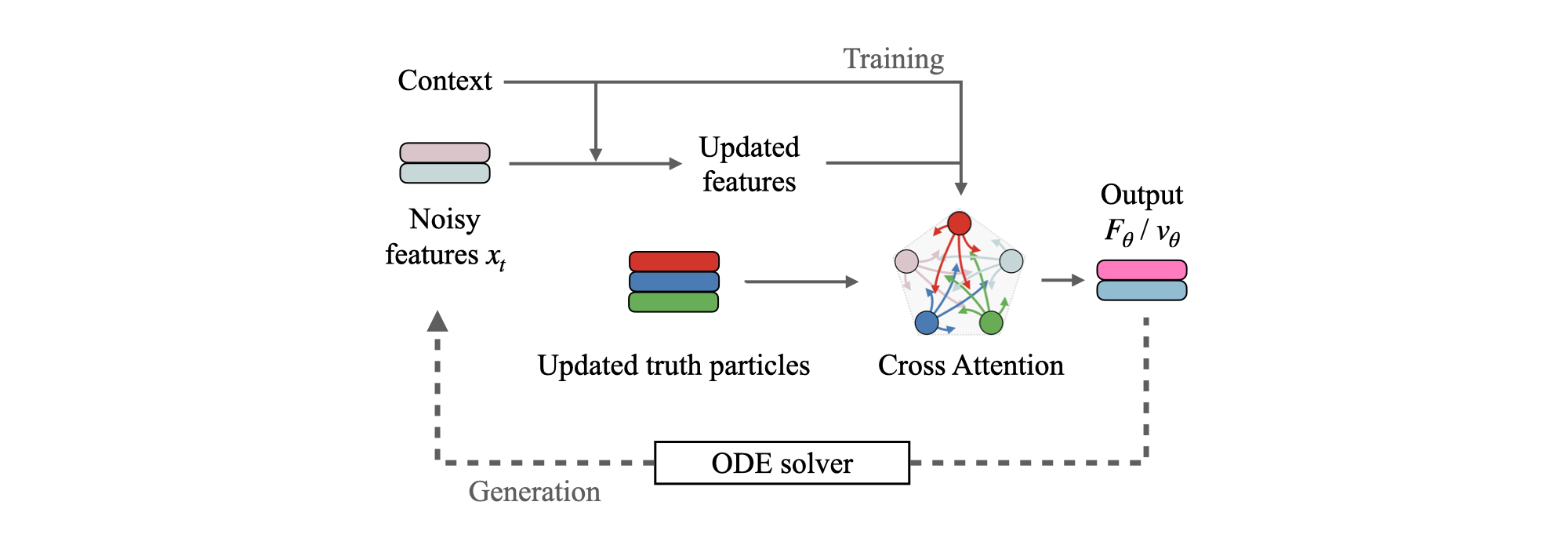}
    \caption{Schematic representation of training and generation processes of the EDM and CFM models. For the training, noise is applied to the particle flow objects. Using the updated truth information from the first network, the denoised properties are obtained through a cross-attention block. For generation, the network iteratively denoises the objects using the ODE solver.}
    \label{fig:diff_model}
\end{figure}
\begin{enumerate}[a)]
    \item Input set Encoding and Cardinality Prediction \\
    This segment is similar to the slot attention approach, as depicted at the top of Fig.~\ref{fig:all_nets}. The set of truth particles characterized by $p_T, \eta, \phi$, and $|q|$ is enriched by a sinusoidal positional encoding, embedded using a multilayer-perceptron (MLP) and subsequently updated through two transformer encoder blocks. An overall embedding $G(T)$ of the truth event $T$ is aggregated using the node-level representation. It is used to predict the categorical distribution of the output set cardinality via an MLP.
    
    \item Feature prediction \\
    This network component takes as input the set of PFOs with noised features $x_t$, combined with self-conditioning features $\tilde{x}/\tilde{v}$ and the current timestep. These features are concatenated with the sinusoidal positional encoding and embedded using an MLP. The embedding process also incorporates the global representation of the truth event $G(T)$, event scaling information, and the sinusoidal embedding of the timestep as contextual information. Then, updated features are processed through three multi-head cross-attention blocks. Each block employs two cross-attention mechanisms to first update the PFOs using the truth particles and then update the truth particles using the PFOs. The final representation is then processed by an MLP together with the contextual information to produce $F_{\theta}$ or $v_{\theta}$. In the diffusion model, the output is combined with a skip connection to generate the denoised estimate $D_{\theta}$. A schematic representation of this process is shown in Fig.~\ref{fig:diff_model}.
    
    \item Inference \\
    Since our goal is to predict the cardinality of the generated set of particles, we split the inference process into two steps. Initially, we predict the cardinality of the generated particles from the set of truth particles and initialize them with Gaussian noise. Subsequently, we employ the PNDM method~\cite{liu_pseudo_2022} in conjunction with our network to iteratively sample the reconstructed particle features from the noise. Following generation, objects with a $\Delta R > 0.5$ with respect to the truth-jet are removed to stabilize against outliers.
\end{enumerate}
\begin{table}[b]
    \centering
    \begin{tabular*}{0.7\linewidth}{@{\extracolsep{\fill}}lr}
    \toprule
    \multicolumn{2}{@{}l}{\textbf{Hyperparameters}}\\\hline
    Batch size & 100\\
    Optimizer & AdamW~\cite{loshchilov2019decoupled}\\
    Weight decay & 0.01\\
    Learning rate & $10^{-4}$\\
    \# of epochs & 150\\
    \# of time steps & 25\\
    Gradient norm clip & 1.0\\\hline
    Max output particles & 51\\
    \# of transformer blocks & 2\\
    \# of cross attention blocks & 3\\
    Trainable parameters & 1,742,519 \\
    \bottomrule
    \end{tabular*}
    \caption{Network hyperparameters of the diffusion models.}
    \label{tab:hyperparams}
\end{table}
\section{Training and Loss}
\label{sec:training}
For the Slot Attention model, we adopt a familiar loss function outlined in Ref.~\cite{fastsim}. The central element involves the application of the Hungarian algorithm~\cite{HUNGARIAN} for particle matching. For each pair of input and output particles from the two sets, we calculate the squared differences for the particle features $\log p_\text{T}$ and $\eta$. Due to the periodicity of $\phi$, we use the cosine loss $2(1-\cos\Delta\phi)$, which recovers the squared difference when Taylor expanded to the second order. Additionally, we include the Binary Cross Entropy (BCE) loss for the charge.

Hungarian matching determines the pairing configuration that minimizes the sum of all pair losses per event. The total loss comprises this sum and the Cross-Entropy (CE) loss for the cardinality prediction. 

Originally, the set approach was chosen to ensure permutation invariance, as no enforced ordering has a physical motivation. However, for the diffusion, we found that due to the sampling, better performance is achieved if ordering is applied so that every particle stays on its own trajectory. Consequently, there is no need for Hungarian matching, and we use regular Mean Squared Error (MSE) loss for the diffusion models.
We also weigh the EDM model's losses as defined in Eq.~\ref{eq:diffloss}.

The training, validation, and test datasets comprise 1,000,000, 20,000, and 300,000 truth events, respectively. The trainings were performed on an NVIDIA RTX A5000.
\begin{figure}[t]
    \centering
    \includegraphics[width=\textwidth]{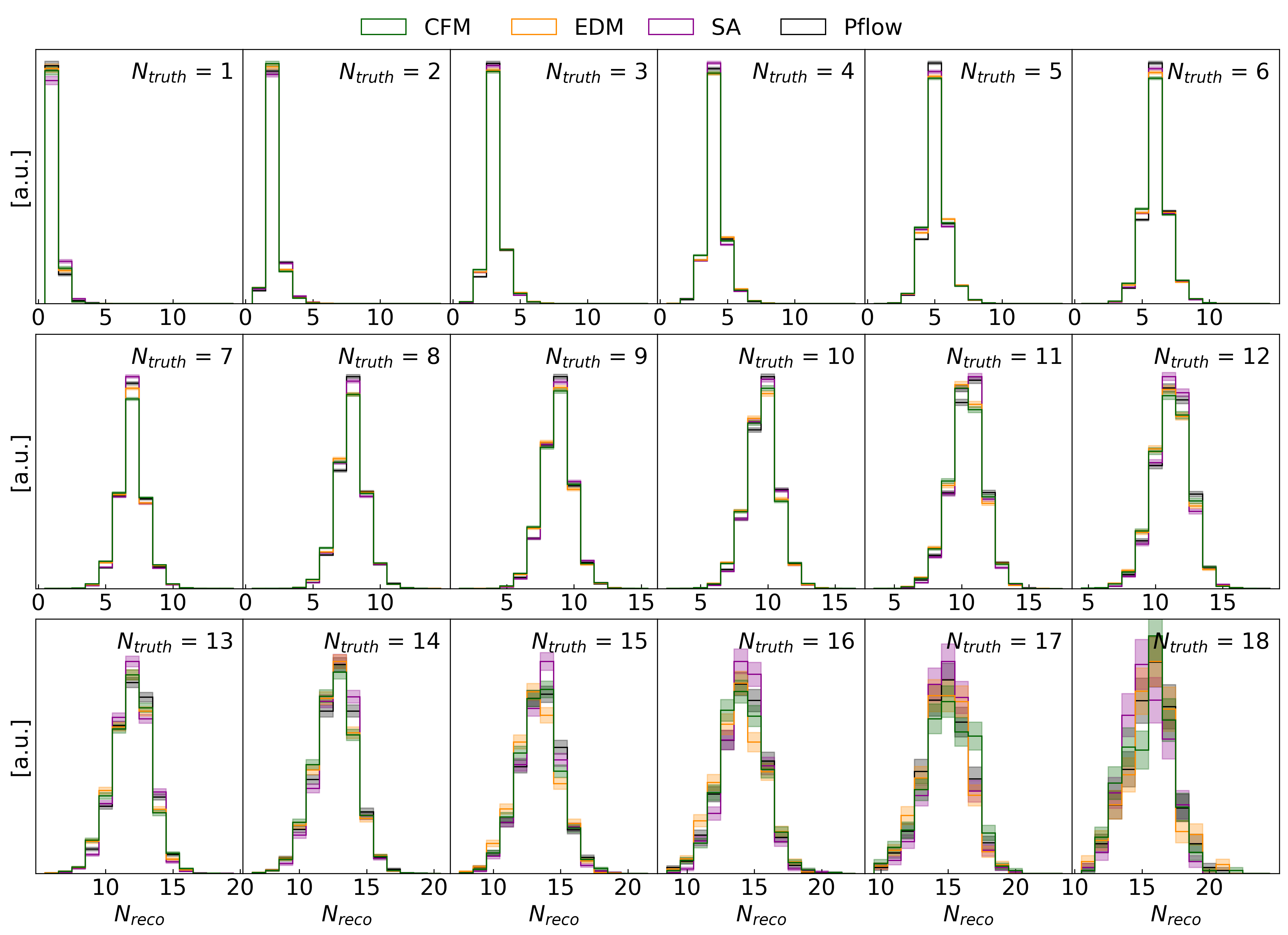}
    \caption{Cardinality distributions for the generated PFOs and target sets for different input set cardinalities.}
    \label{fig:cardinality}
\end{figure}
\section{Results}
\label{sec:results}
Given the task of set generation, we can assess multiple aspects of network performance. First, we analyze properties of the set as a whole, such as set cardinality and inclusive particle distributions. Second, we examine the properties of individual set constituents. Additionally, for our application, we aim to study resolution modeling, which entails examining the variation in generated particle properties given the same input values. These aspects are discussed separately in the following subsections.
\subsection{Set-based Performance}
\begin{figure*}[t]
    \centering
    \begin{subfigure}{0.325\textwidth}
        \centering
        \includegraphics[width=\textwidth]{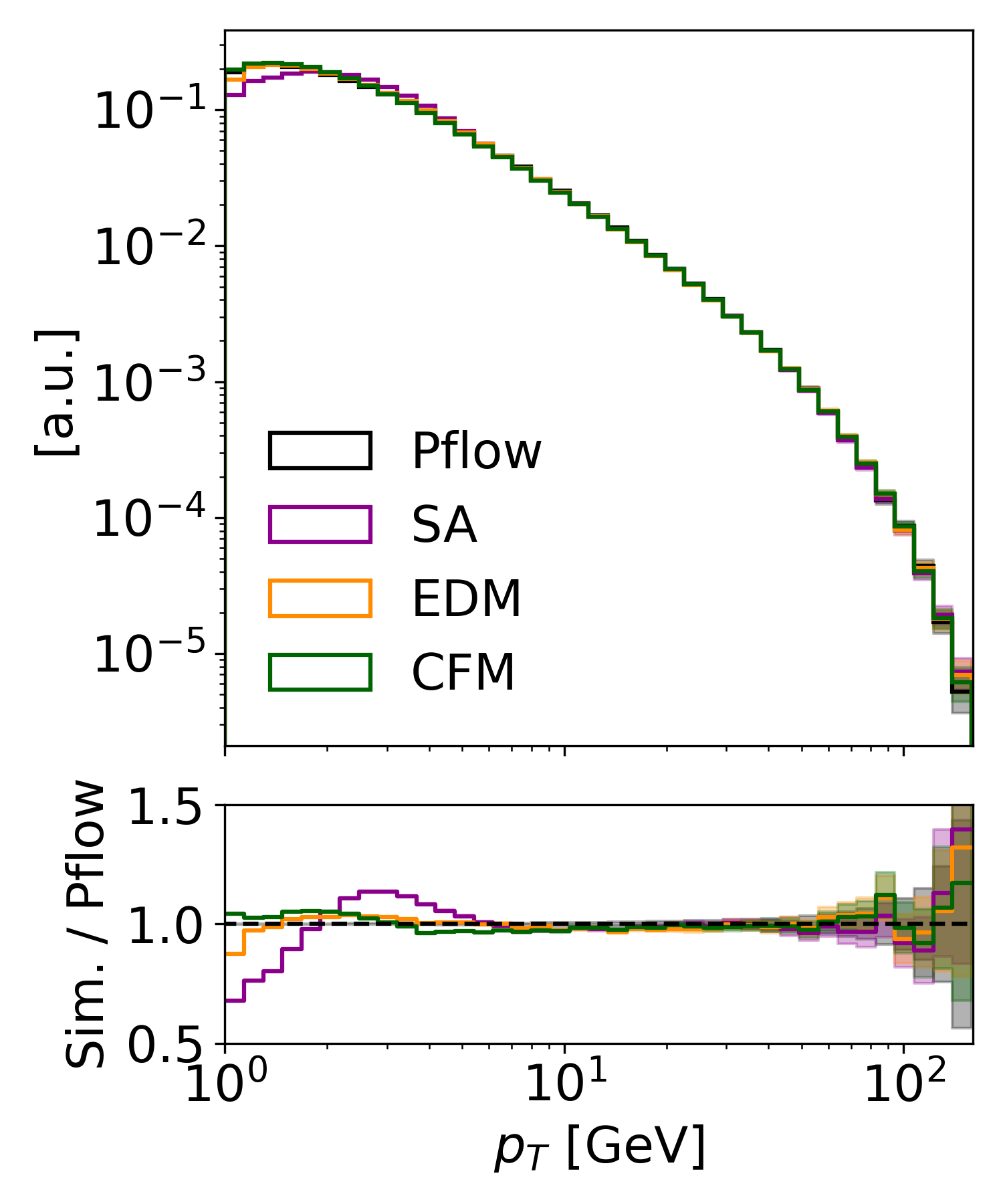}
    \end{subfigure}
    \begin{subfigure}{0.325\textwidth}
        \centering
        \includegraphics[width=\textwidth]{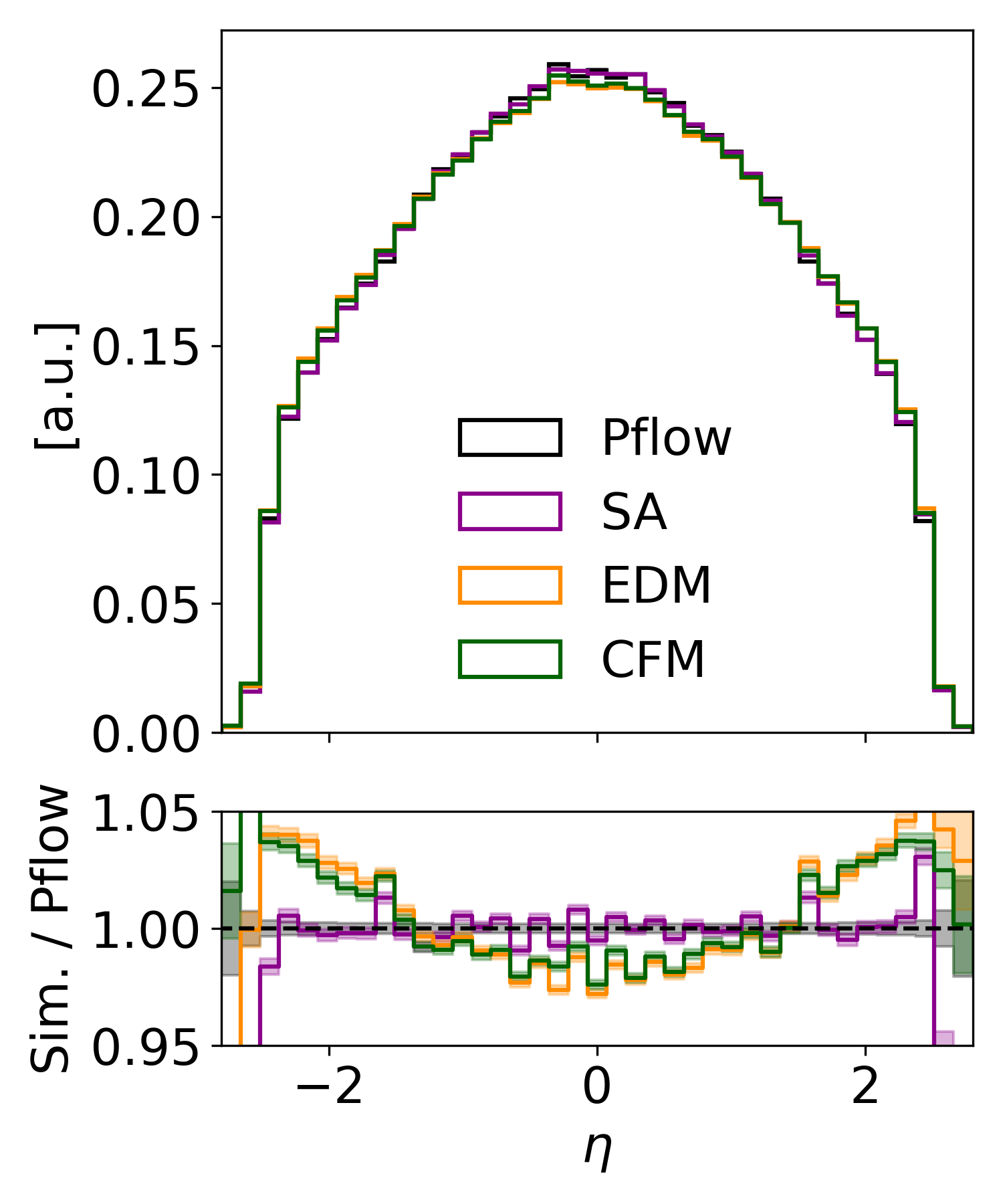}
    \end{subfigure}
    \begin{subfigure}{0.325\textwidth}
        \centering
        \includegraphics[width=\textwidth]{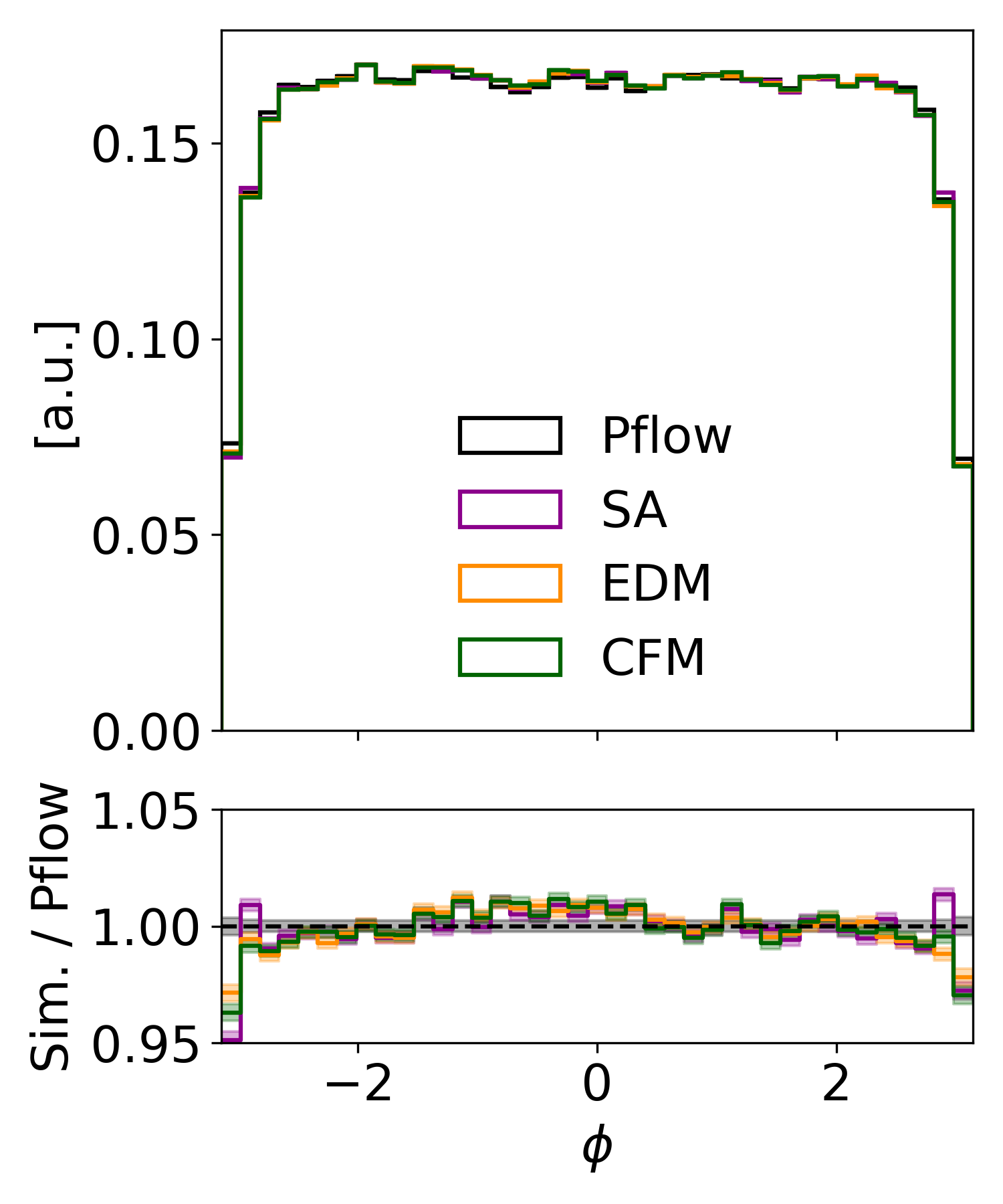}
    \end{subfigure}
    \caption{Generated and target distributions of $p_T$, $\eta$ and $\phi$ for all particles in the test set.}
    \label{fig:1dplots}
\end{figure*}

The cardinality prediction network follows the same principles and structure for all used approaches; hence, similar results are expected. Minor deviations may occur as the networks are trained simultaneously with the property prediction network using combined losses. Figure~\ref{fig:cardinality} compares the generated and target set cardinalities for different input set cardinalities. The distributions for low cardinalities peak at the truth cardinality, with the width of the distribution increasing as the number of truth particles rises. Higher particle counts in the event lead to more instances of misreconstructions and inefficiencies. For very high cardinalities, the peak position is lower than the truth cardinality since dense environments are more prone to reconstructing several neutral particles as one. Both networks closely match the target distributions, even for high cardinalities where the event count is low, with predictions aligning with targets within statistical uncertainties.
Next, we examine the marginal distributions of the set constituents depicted in Fig.~\ref{fig:1dplots}. These histograms encompass all particles from all events in the test set. The diffusion models outperform the baseline model in the $p_T$ distributions for low energies. Both models perform well in the intermediate and high $p_T$ range. The Slot Attention model slightly outperforms the diffusion models in the distribution of $\eta$. The bias comes from the fact that low $p_T$ particles tend to have higher $|\eta|$ and vice versa.
 However, the mismatch results in a deviation of a few percent, which is not deemed very significant. The distributions of $\phi$ closely match the target distribution within statistical errors for both networks. 
\begin{figure}[b]
    \centering
    \includegraphics[width=0.75\textwidth]{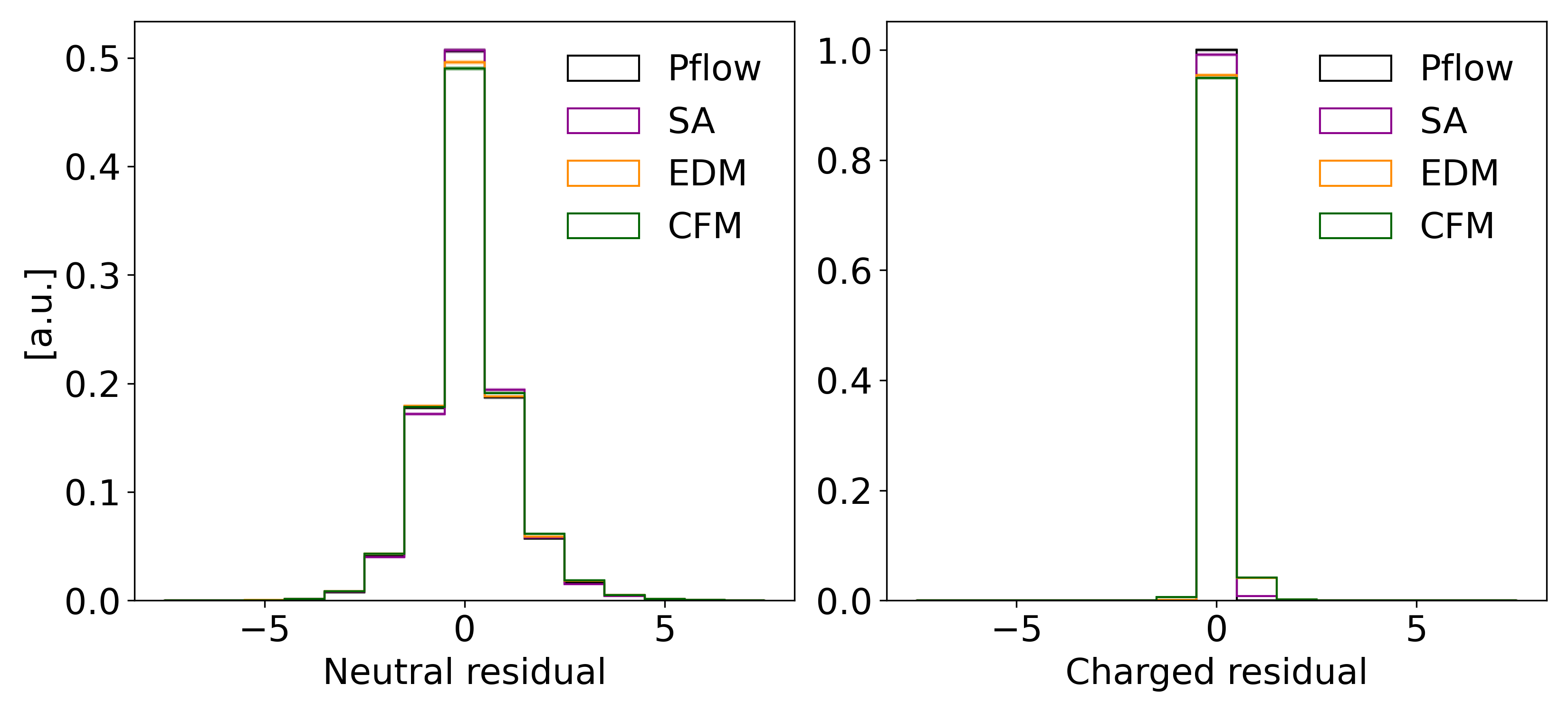}
    \caption{Difference of truth and reconstructed cardinality per event split by neutral and charged constituents.}
    \label{fig:class_pred}
\end{figure}

The feature prediction task involves classifying particles as charged or neutral. Given the initially predicted total cardinality, the network must learn the number of charged and neutral constituents in the set. Since the total cardinality prediction displays good agreement, we can separately compare the neutral and charged contributions to analyze class prediction performance. Figure~\ref{fig:class_pred} illustrates the difference between the number of truth and reconstructed particles per event, separated for neutral and charged constituents. The target distribution for charged particles is a delta function since track presence ensures perfect classification. However, the reconstruction of neutral particles is more prone to inefficiencies and fakes, resulting in a wider distribution. Slot Attention demonstrates excellent agreement with the target, while the diffusion models exhibit some misclassification, as achieving binary prediction accuracy is challenging in diffusion models. 

Overall, we conclude that both models are able to reproduce the salient properteis of the sets of PFOs.
\begin{figure*}[b]
    \centering
    \begin{subfigure}{0.328\textwidth}
        \centering
        \includegraphics[width=\textwidth]{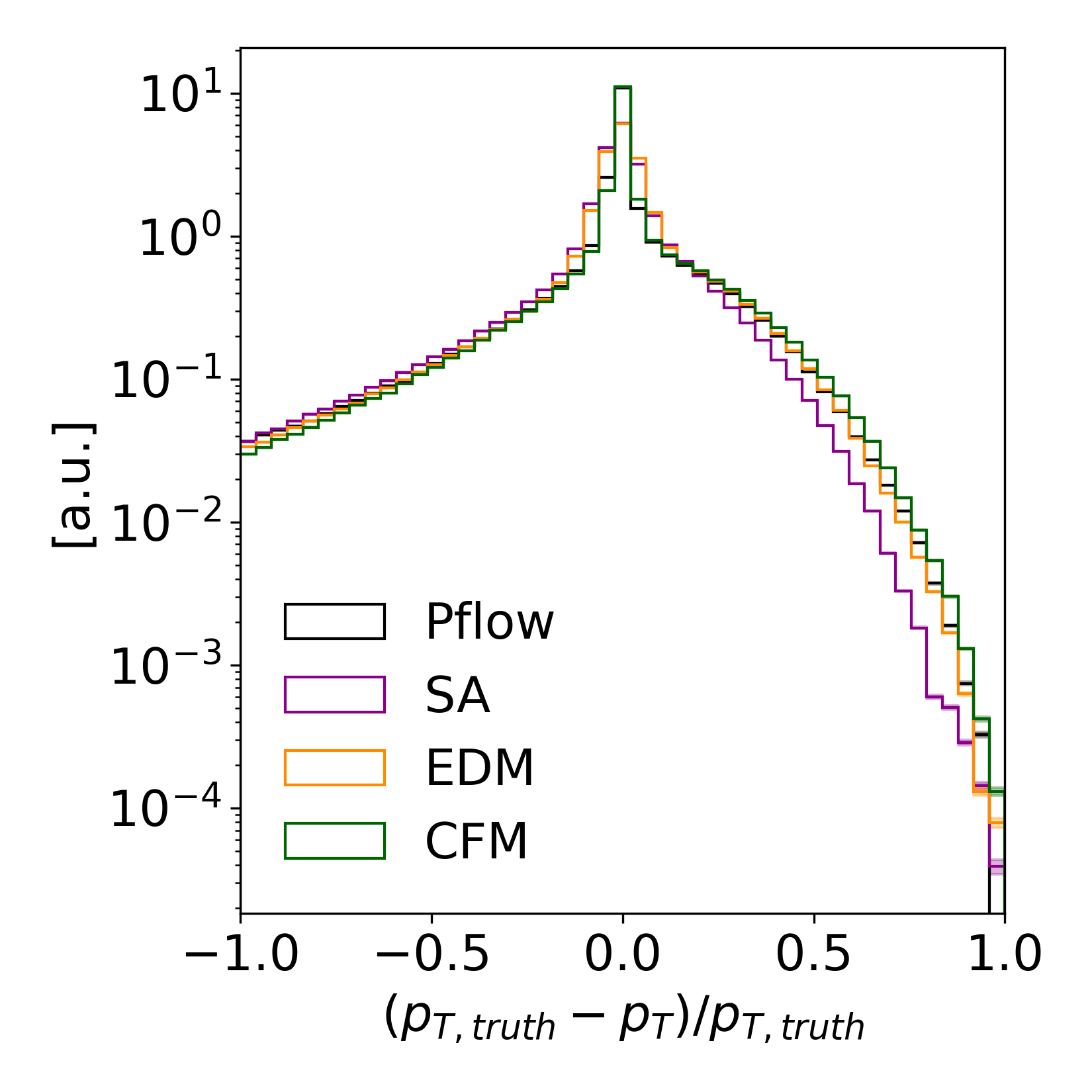}
    \end{subfigure}
    \begin{subfigure}{0.328\textwidth}
        \centering
        \includegraphics[width=\textwidth]{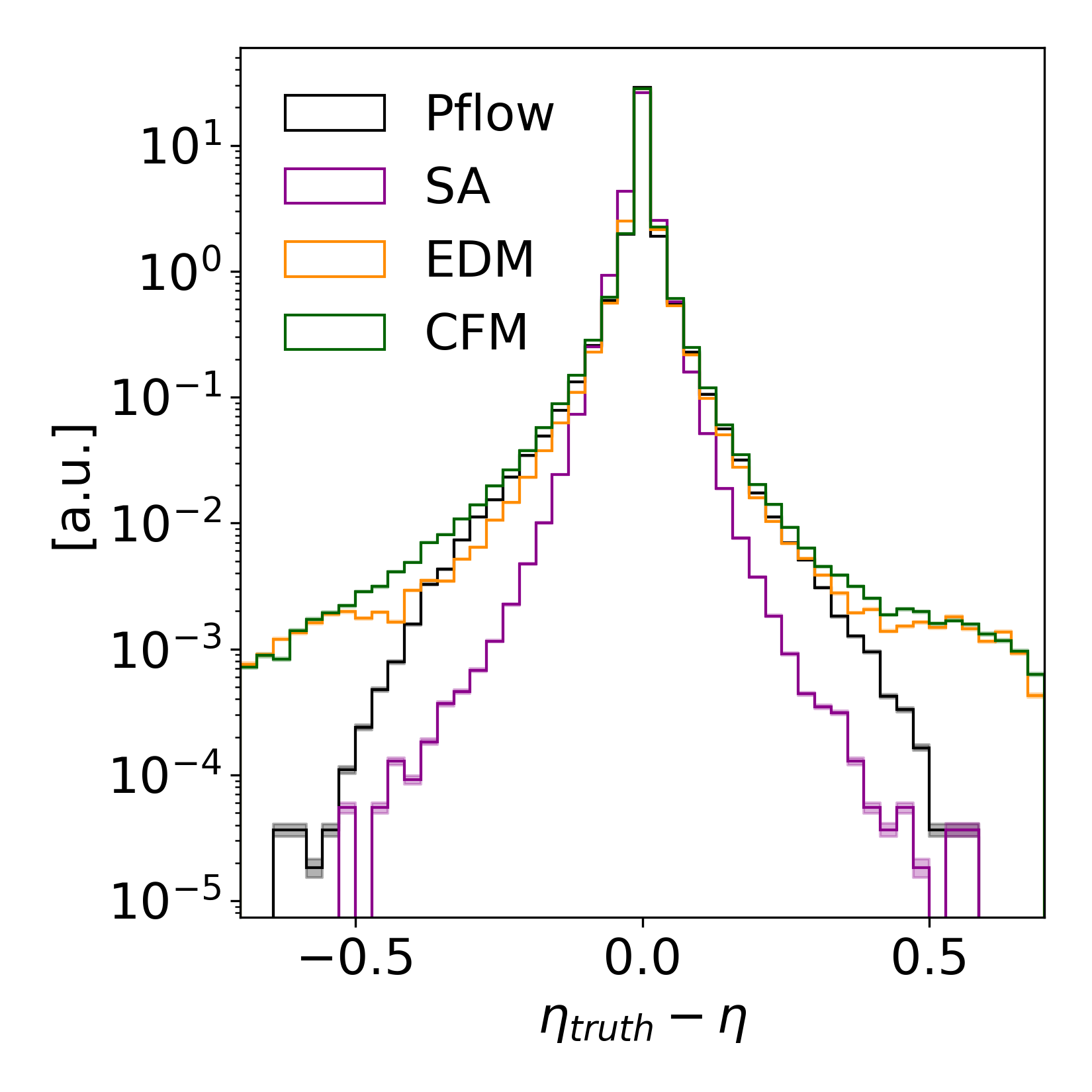}
    \end{subfigure}
    \begin{subfigure}{0.328\textwidth}
        \centering
        \includegraphics[width=\textwidth]{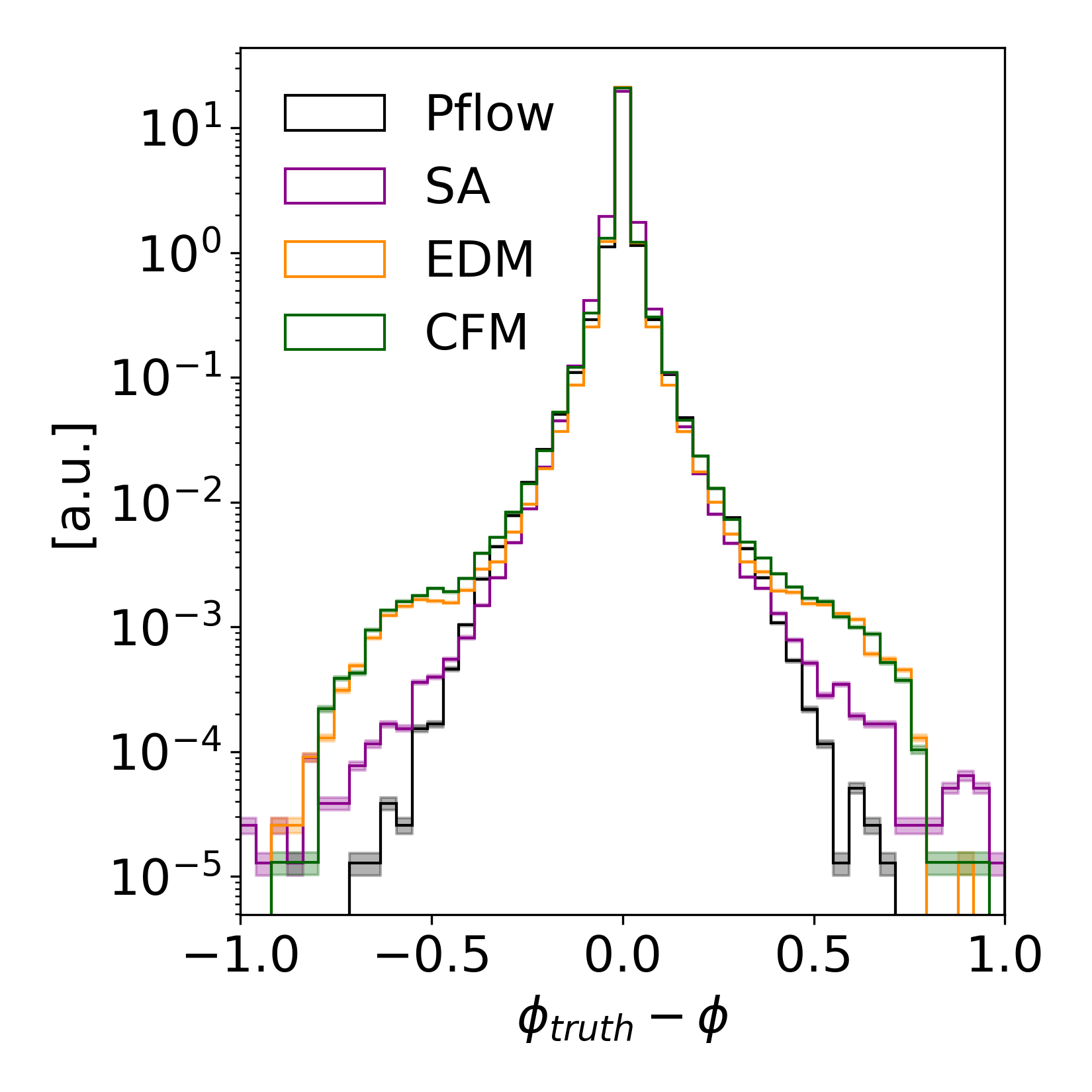}
    \end{subfigure}
    \caption{Residual distributions of $p_T$, $\eta$ and $\phi$ for particles matched between ground truth particles and target/fast simulated reconstructed particles.}
    \label{fig:1d_residuals}
\end{figure*}
\subsection{Constituent-based Performance}
A more nuanced investigation involves a per-event comparison of set constituents. While Fig.~\ref{fig:1dplots} illustrates that \textit{on average}, the generated particles have correct features, mismatches among certain features may occur within an event, which cannot be captured in this plot. To explore this further, we conduct Hungarian matching using the same configuration as during training. However, here, we match the truth set with the output set. With the matched pairs, we examine their differences in $p_T$, $\eta$, and $\phi$.

These residual distributions are depicted in Fig.~\ref{fig:1d_residuals}. For angular quantities, the diffusion models perform better than Slot Attention in matching the peak of the distribution but exhibits wider tails originating from outliers, particularly in $\eta$. Conversely, Slot Attention displays a narrower distribution in $\eta$, suggesting that generated particles are not sufficiently smeared with respect to the truth. This trend is also evident in the right tail of the $p_T$ distribution. Here, the diffusion models demonstrate better agreement. However, EDM deviates from the target distribution in the peak area similarly to the SA, while CFM has a very good match. Notably, this area is primarily influenced by low-$p_T$ particles, which are challenging to model. 
\begin{figure}[t]
    \centering
    \includegraphics[width=\textwidth]{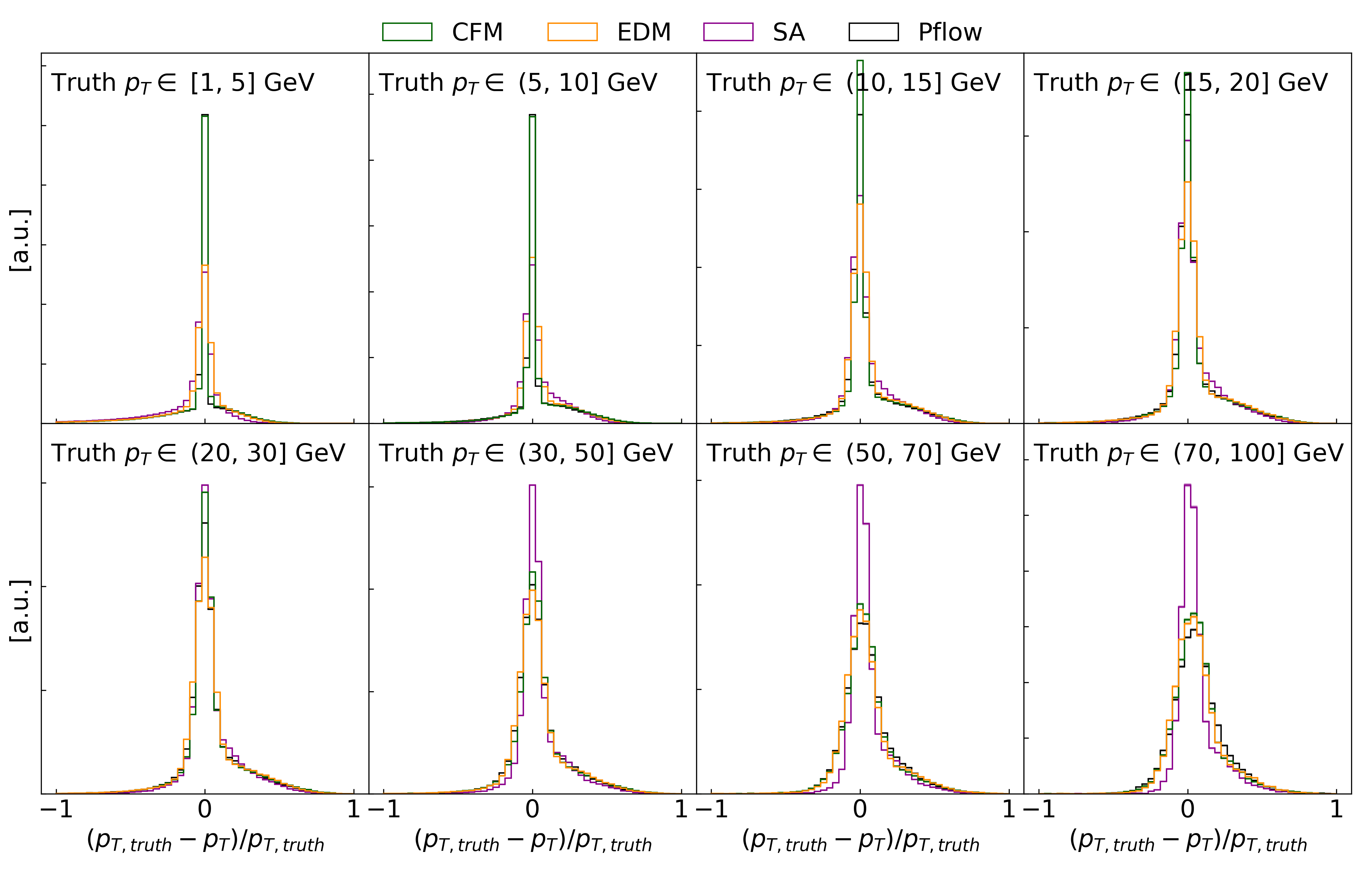}
    \caption{Residual distributions of $p_T$ for matched particles split for different truth $p_T$.  }
    \label{fig:res_split_pt}
\end{figure}
\begin{figure}[b!]
    \centering
    \includegraphics[width=0.6\textwidth]{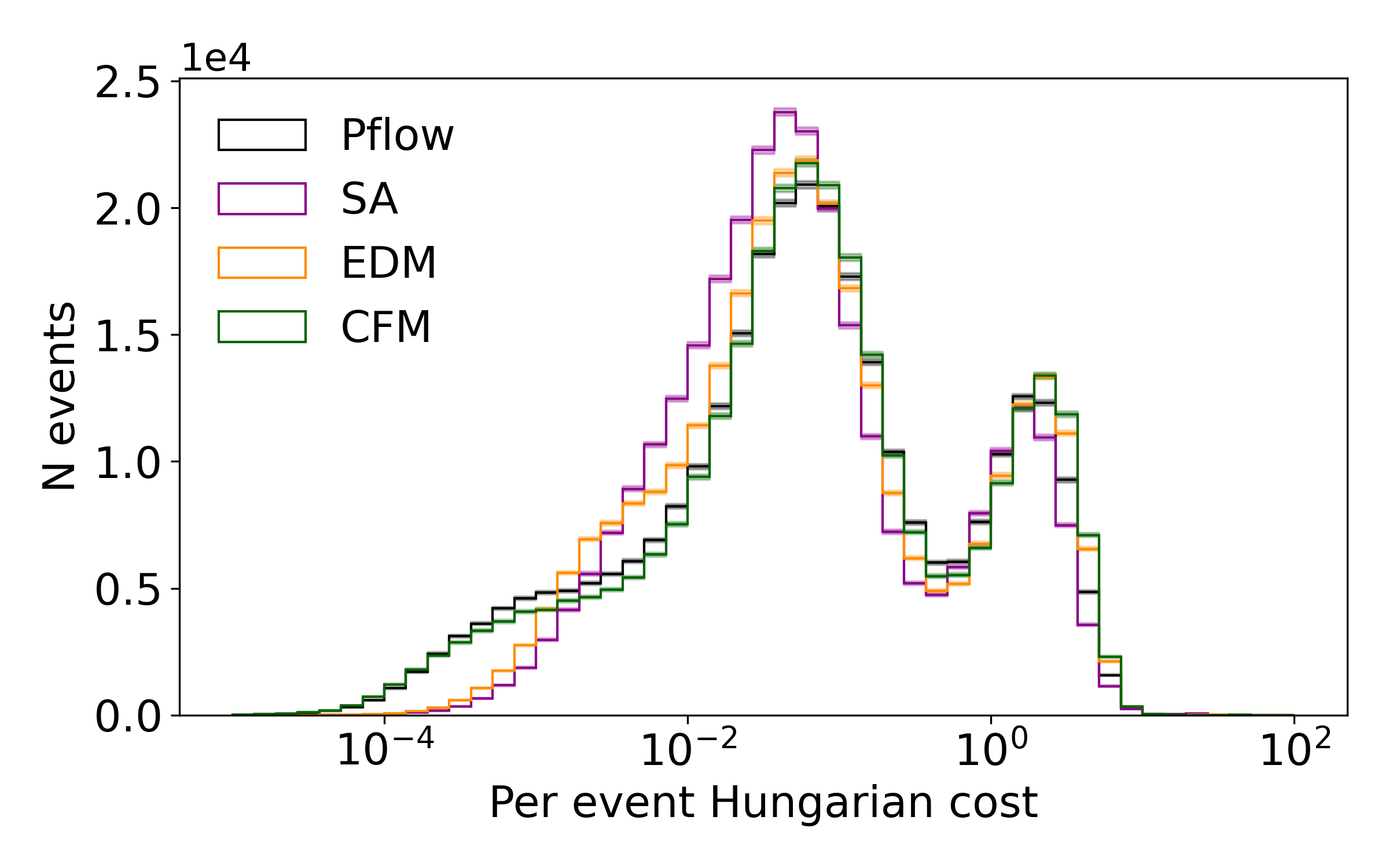}
    \caption{Per event Hungarian cost for truth matching normalized by number of matched particles in the event.}
    \label{fig:hungarian}
\end{figure}

To gain deeper insights, we depict the residual $p_T$ distribution for different bins of the matched truth particle $p_T$ in Fig.~\ref{fig:res_split_pt}. Initially focusing on the target distributions, we observe that low-energy particles form a narrow distribution with sharply declining tails comprised of mismatched or noisy particles. As the truth $p_T$ increases, the distributions widen,
likely driven by worsening track momentum resolution. All tend to overestimate the residuals for low energies. For higher energies, we observe excellent agreement for the diffusion models, while slot attention predicts distributions that are too narrow, underestimating smearing effects and predicting values too close to the truth.

We can also assess constituent-based performance at the event level. Similar to the procedure in the training loss, we can extract the Hungarian cost for each event from the truth and reconstructed particle features, which can conveniently be used as a summary metric. It is depicted in Fig.~\ref{fig:hungarian}. Intuitively, this plot can be understood as follows: If the generated distribution is shifted to the left of the target, i.e., the Hungarian cost is too low, the predictions are too close to the truth and smearing effects are not well modeled. Conversely, a shift to the right indicates excessive smearing of truth quantities and possibly outliers. The double-peak structure originates from events with and without high $p_T$ particles.

We can summarize previously made observations: Slot Attention exhibits a clear left-shift in the bulk of the data, indicating overly precise predictions. Both networks fail to match the left tail of the distribution, which consists of events with very low energies. In the right peak, we observe a slight shift to the right for the diffusion models. While the diffusion models shows room for improvement on the edges of the distributions, they match the bulk of the data very well and clearly outperform the baseline model. The CFM seems somewhat superior to the EDM, in particular for the per-particle Hungarian cost in Fig.~\ref{fig:hungarian}.

Acknowledging that matching-based comparison is the easiest method for evaluating the performance of constituent modeling, it's important to note a few flaws that can potentially introduce biases. Firstly, particles without a match are not included in the comparison, and since the matching is performed with the truth, a substantial number of particles can be missed, as evident from Fig.~\ref{fig:cardinality}. Additionally, in some cases, a generated event may have the same number of particles as the truth, but one particle was not reconstructed, while a particle from noise with very different features was created. This scenario can occur in reality, but the matching would pair the unreconstructed truth particle with the fake one, resulting in a high Hungarian cost. We also note that when summing the costs of $p_T$, $\eta$, $\phi$ and class we weigh them so every variable on average has approximately the same contribution. Changing the balance would yield different matching results and can influence the performance metrics.

Despite these flaws, matching-based comparison still serves as a useful metric, as these effects will equally occur for both targets and predictions as long as the cardinality is modeled correctly.
\subsection{Resolution Performance}

\begin{figure}[t]
    \centering
        \begin{subfigure}{0.49\textwidth}
        \centering
            \includegraphics[width=\textwidth]{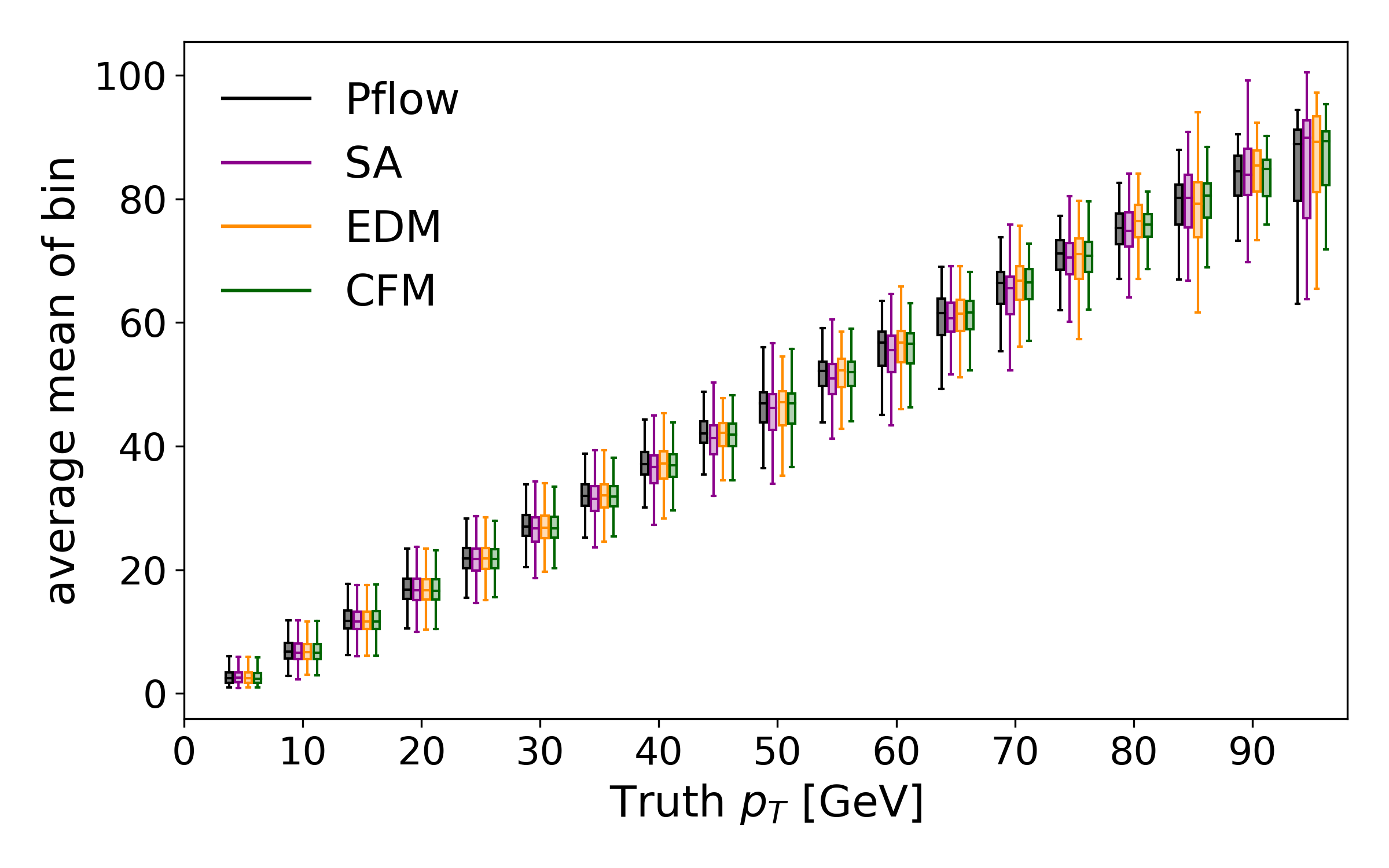}
        \end{subfigure}
        \begin{subfigure}{0.49\textwidth}
        \centering
            \includegraphics[width=\textwidth]{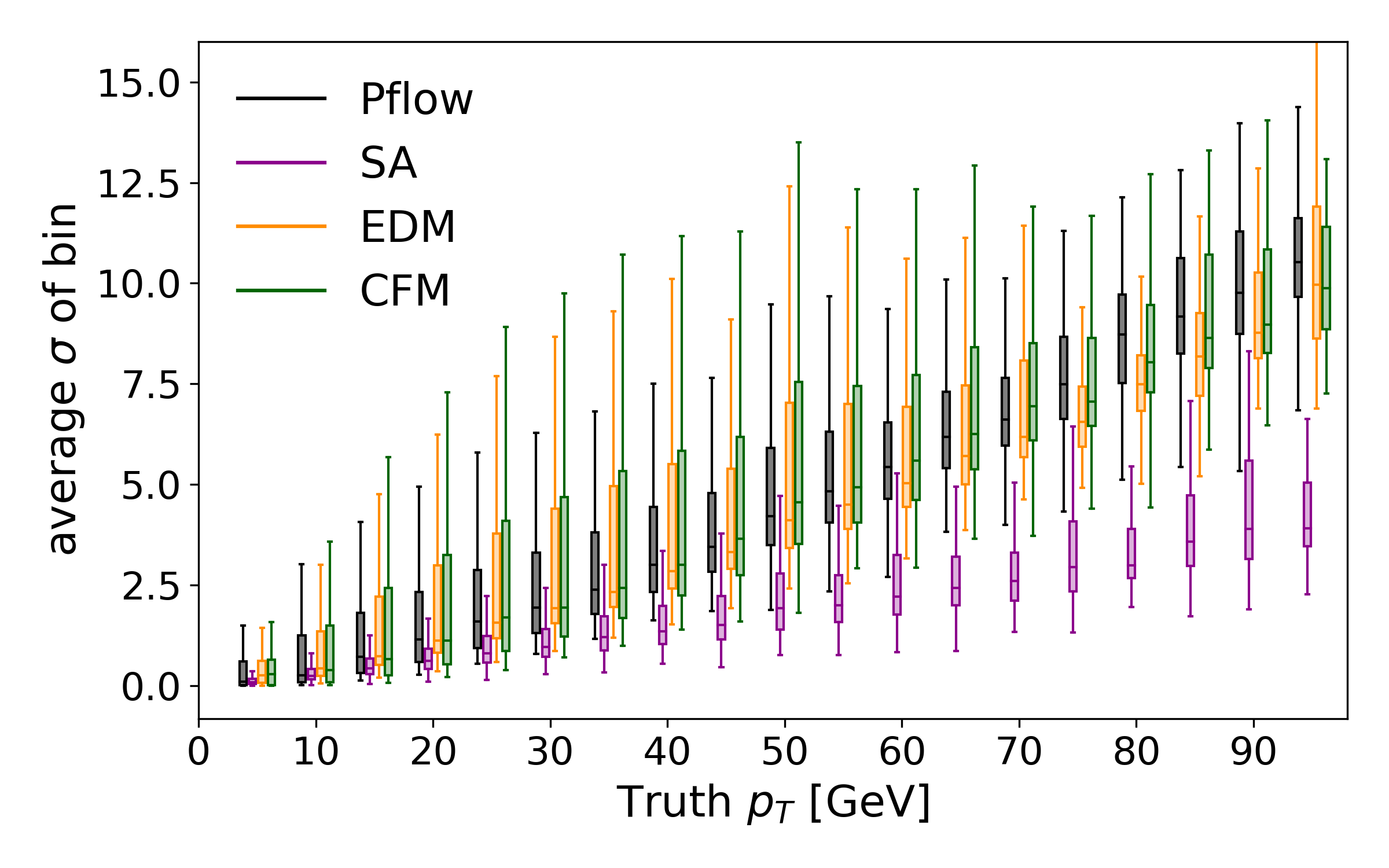}
        \end{subfigure}
    \caption{Means and sigmas of the per-particle replica $p_T$-distributions binned by the matched truth particle $p_T$.}
    \label{fig:mean_sigma}
\end{figure}

A crucial aspect for a good detector simulation is accurate resolution modeling. The stochastic nature of particle-material interactions leads to non-trivial smearing of truth particle features. Conducting detector simulation and reconstruction multiple times on the same truth event results in different outcomes each time. To evaluate how well our model reproduces this variability, we introduce a \textit{replica} dataset, similarly to the concept introduced in Ref.~\cite{fastsim}. We generate 10,000 unique truth events, and for each truth event, we run detector simulation and reconstruction 100 times. After performing truth matching for all replicas, each truth particle has a matched distribution of particles. For each truth particle, we compute the mean and standard deviation of the matched distribution as a proxy for the detector resolution. This is summarized in bins of truth $p_T$ in Fig.~\ref{fig:mean_sigma}. 
\begin{figure}[t]
    \centering
    \includegraphics[width=\textwidth]{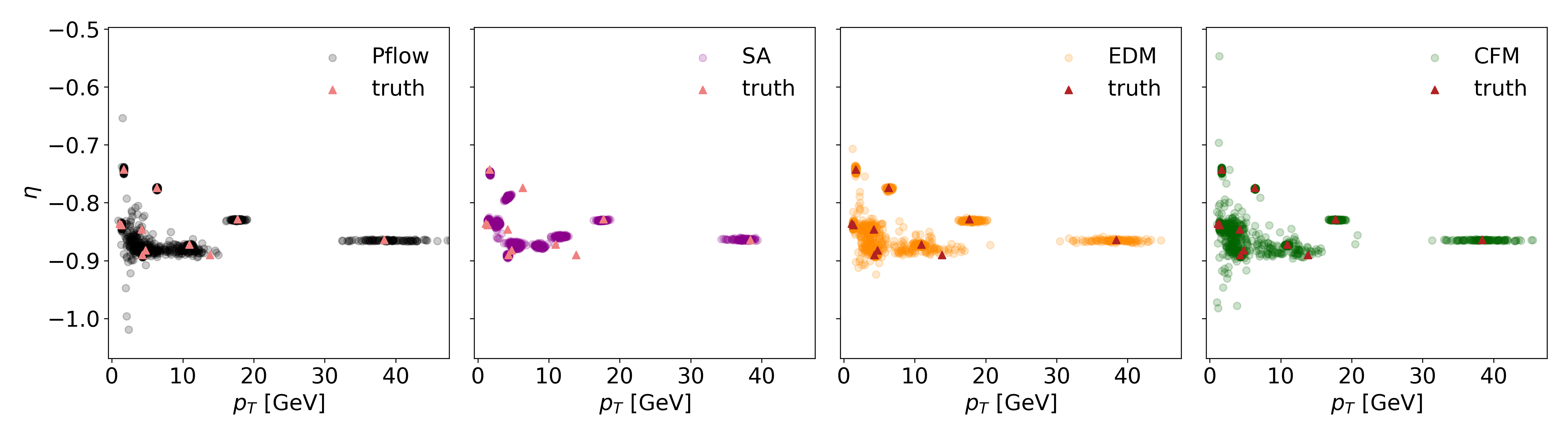}
    \caption{Event display with replicas. The detector simulation is run 100 times for the displayed truth event.}
    \label{fig:ED}
\end{figure}
While the means are modeled comparably well by both models, Slot Attention largely underestimates the standard deviations and therefore doesn't accurately model the resolution. 
However, the diffusion models match the target resolution significantly better. EDM features a slight trend towards underestimating the resolution.

We note that utilizing the replica dataset for training with an extended loss function, as demonstrated in Ref.~\cite{fastsim}, is an option. In such a scenario, the Slot Attention model would yield better performance in resolution modelling. However, the diffusion models consistently outperform it. Since we found no significant advantage for the latter, we opted for the simpler setup without replicas.

For demonstrative purposes, we present an event display with replicas in Fig.~\ref{fig:ED}. The showcased event was selected because it contains particles across a broad $p_T$ spectrum without exceeding a count of particles that would classify it as a rare event. The event display aligns with our previous discussions. Slot Attention fails to accurately model the distribution and does not replicate the low-energy noise. In contrast, the diffusion models exhibit a much closer alignment with the target distribution but are prone to a few noisy outliers.

\section{Conclusion}
\label{sec:conclusions}
In this study, we have advanced a previously introduced approach for fast simulation of particles within dense environments such as jets. Our method involves directly generating reconstructed particles from input truth particles, effectively replacing the conventional processes of detector simulation and reconstruction in a single step.

Central to our approach is the introduction of the diffusion models, which leverage cutting-edge diffusion techniques on graph-valued data. We conducted a comprehensive analysis of the networks' performance, focusing on three key aspects: overall set quantities, features of individual set constituents, and resolution modeling, a significant challenge in detector simulation. Both diffusion models outperform slot attention with respect to resolution modelling without the need for a training dataset containing multiple detector replicas of the same underlying truth event. Overall, the CFM model outperformed the EDM, seen most clearly for the per-event Hungarian matching cost (Fig.~\ref{fig:hungarian}).

To accurately assess resolution performance, we introduced the replica dataset, enabling us to evaluate generation outcomes multiple times per ground truth. Our findings demonstrate that the diffusion models significantly outperform the baseline model in resolution modelling while exhibiting comparable performance in overall set constituent features. 

Looking ahead, our ultimate objective is to develop a fast simulation package that is universally applicable across various events. Future directions could involve scaling up the model to process full events and, therefore, handle higher cardinalities, although the performance implications of such scaling remain uncertain, particularly for events with several hundred particles. Alternatively, implementing event partitioning could address the cardinality issue, but would require training on a diverse range of detector signatures, including jets of varying energy ranges and isolated objects. Using the results of this paper, Ref.~\cite{letter} makes progress on some of these challenges towards a full end-to-end approach.

\section*{Data and Code Availability}

The code for this paper can be found at \url{https://github.com/dkobylianskii/fastsim-advanced}. The COCOA dataset can be found at Zenodo at \url{https://zenodo.org/records/11383444}.

\section*{Acknowledgement}
We thank Yaron Lipman and Vinicius Mikuni for fruitful discussions. ED, EG, NK, DK, and NS are supported by the BSF-NSF grant 2028 and the ISF Research Center 494. BN is supported by the U.S. Department of Energy (DOE), Office of Science under contract DE-AC02-05CH11231.

\newpage

\bibliography{fastsim}
\end{document}